\documentclass[letterpaper,twocolumn,10pt]{article}
\pdfoutput=1

\usepackage{cite}

\usepackage{graphicx}

\graphicspath{{./data/},{./figures/}}
\DeclareGraphicsExtensions{.pdf,.png,.jpg}

\usepackage{fullpage}

\usepackage{array}
\usepackage[T1]{fontenc}
\usepackage[utf8]{inputenc}
\usepackage[english]{babel}
\usepackage{textcomp} 

\usepackage[usenames,dvipsnames,svgnames,table]{xcolor}

\usepackage{booktabs}

\usepackage{amsmath}

\usepackage{comment}

\usepackage{placeins}

\newcommand{\todo}[1]{}

\usepackage[hyphens]{url}
\usepackage[hidelinks]{hyperref}

\newcommand{\wattsup}{\textsc{Watts Up?\ pro}}

\newcommand{\etal}[1]{et al.~\cite{#1}}

\providecommand{\tightlist}{%
  \setlength{\itemsep}{0pt}\setlength{\parskip}{0pt}}

\hypersetup{pdfauthor={%
    Eddie Antonio Santos, Carson McLean, Christopher
    Solinas, Abram Hindle
  },
  pdftitle={%
        How does Docker affect energy consumption?
        Evaluating workloads in and out of Docker containers
  }
}

\begin{document}

\title{%
  How does Docker affect energy consumption?\\
  Evaluating workloads in and out of Docker containers
}

\date{}

\author{
  Eddie Antonio Santos\thanks{%
    Department of Computing Science, University of Alberta, Edmonton, Canada
  }
  \and
  Carson McLean\footnotemark[1]
  \and
  Christopher Solinas\footnotemark[1]
  \and
  Abram Hindle\footnotemark[1]
}

\maketitle

\subsection*{Abstract}

\textbf{Context:} Virtual machines provide isolation of services at the cost of hypervisors and more resource usage.
This spurred the growth of systems like Docker that enable
single hosts to isolate several applications, similar to VMs, within a low-overhead abstraction called
\emph{containers}.

\noindent \textbf{Motivation:} Although containers tout low overhead
performance, do they still have low energy consumption?

\noindent \textbf{Methodology:} This work
statistically compares ($t$-test, Wilcoxon) the energy consumption of three application workloads in Docker and on bare-metal Linux.

\noindent \textbf{Results:} In all cases, there was a statistically
significant ($t$-test and Wilcoxon $p < 0.05$) increase in energy
consumption when running tests in Docker, mostly due to the
performance of I/O system calls.

\paragraph{Keywords:} virtualization, docker, containerization, energy
consumption, cloud computing, microservice.

\section{Introduction}
\label{sec:introduction}

\emph{Virtualization} provides a number of benefits when deploying
software, such as process isolation and resource control.
Process \textbf{isolation} means that software developers can make strong
assumptions about the state of the system, including the operating system
configuration, and having the exact software dependencies needed for the
system.
Virtualization often allows for \textbf{resource control} such that
operators can configure precisely how much CPU, memory, or access to network
interfaces a particular application has. Virtualization platforms often use
\emph{images}, snapshots of the complete system needed to run an application,
thus deploying an app is as easy as instantiating an image.
Traditionally, virtualization has been implemented through \emph{virtual
machines}, in which one machine may host several guest operating systems.
However, the intervention of the hypervisor,\footnote{%
    We use the term hypervisor for any virtual machine monitor that is hosted
    on top of an existing operating system, or is a module of the host
    operating system kernel, such as KVM~\cite{KVM}.
} means that applications effectively must use two kernels---directly through
the guest operating system, and indirectly through the hypervisor---when
accessing resources such as network and storage. This may be considered an
undesirable overhead.
This prompted the need for a low overhead virtual machine.
Recently, sophisticated features in the Linux
kernel---namely, namespaces and control groups---made a new form of
low overhead virtualization possible: \emph{containerization}.
Containers are a lightweight alternative to virtual machines, as they offer
isolation (processes, file-system, network) and resource control (CPU,
memory, disk) without the overhead of an additional
kernel.
Container management software such as Docker\cite{what_docker}, LXC~\cite{LXC}, and
\texttt{rkt}~\cite{rkt}, are quickly displacing virtual machines as the
virtualization solution of choice~\cite{docker_adoption,clusterhq_survey}.

Given
the blistering pace of the adoption of containerization, what is the
impact of containerization on energy consumption?
Changes in software have significant and
measurable differences in power and energy
consumption~\cite{hindle_green_2012,zhang_accurate_2010,ellis_case_1999,vasic_making_2009,gupta_detecting_2011}.
Since containerization, in principle, lacks the overhead of virtual machines,
clearly it should consume a similar amount of energy as a bare-metal
configuration.

In this paper, we empirically test this assumption against numerous
measured workloads, run with and without containerization.
In practice,
container providers such as Docker \emph{do} add additional overheads, such as
the AUFS file system, and an abstracted networking layer. We seek to quantify
the impact that these overheads have on energy efficiency.  We compare
the energy consumption of various scenarios run on bare-metal Linux---that is,
the applications are running on one kernel, without any virtualization at
all---in contrast to Docker-managed containers, using ``off-the-shelf'' Docker
images. We use total system power consumption (or ``wall power'') to
estimate total energy consumption. We run several iterations of each
experiment, the results of which we present and explain why we see differences
in energy consumption between bare-metal Linux and Docker.

This work suggests that there is no free lunch for containerization in terms of
energy consumption. Containerization implies a trade-off between energy and
maintainability, and it is up to the individuals or teams in charge of deployment
to determine which is more costly in their particular scenario.

\section{Prior Work}
\label{sec:related-work}
\label{sec:prior-work}

Previous work has focused on \emph{virtual machine} power and energy
consumption.
Xu~\etal{xu_energy_2015}
measured CPU and total power usage in both Xen and KVM hypervisors. They found
that Xen generally has a greater power overhead than KVM when processing
network traffic, attributed to ``excessive interrupt requests''.
They found that as the load is more evenly distributed among
virtual machines, power consumption increases. This paper elaborates on
the effect of Docker on network energy consumption.

Some work has compared virtual machines to containers directly.
Morabito~\cite{morabito_power_2015} compared the power usage of traditional
virtual machine hypervisors (KVM, Xen) to container based virtualization
(Docker, LXC). In all cases, the container style virtualization used
marginally less power, but overall neither virtualization method showed
significant difference. Morabito did not consider runtime differences, hence
this work cannot make conclusions about overall energy consumption.  Further, there
was no comparison to bare-metal Linux performance. Both of these concerns are
addressed in our work.
Van Kessel~\etal{van2016power} used internal hardware sensors to determine the
difference in power consumption of Xen against Docker. They found that
Docker is more efficient on CPU-bound and disk bound loads. In contrast, our
work compares against bare-metal Linux measuring wall power instead of
internal power sensors to quantify the abstractions provided
by Docker.
Shea~\etal{shea_power_2014} compared the power consumption of network transactions
using virtualization such as KVM, Xen, and OpenVZ, in contrast to a bare-metal
system. Only OpenVZ can be considered container-based virtualization. They
measured both wall power and CPU power using Intel's Running Average Power
Limit  (RAPL). The authors found that power measured through RAPL was
always a fraction of the measured wall power. They found a difference in the
power overheads of network transactions on different virtualization platforms.
However, they concluded that the overheads were tunable.
Our work concentrates only on Docker's container-based virtualization.  We
measure wall power only, because we wanted to capture the total system power
usage. Additionally, we measured more scenarios than just network transactions.

Other work has evaluated container performance metrics such as run time, CPU
usage, and network utilization.
Felter~\etal{felter_updated_2015} compared CPU, memory, I/O, and network
performance of Docker and KVM against bare-metal Linux. In most cases, Docker
adds little overhead, and almost always outperforms KVM\@.
They also tried sample loads on Redis and MySQL\@. They found that, in some
cases such as the Redis example, Docker performs comparably to bare-metal when
configured appropriately.  The authors found that Docker's UnionFS file system
abstraction has negative performance penalties compared to a standard Linux
file system. In contrast, our work directly measures energy consumption of
running similar benchmarks, both on bare-metal Linux compared to within a
Docker container.
In general, quicker runtime is correlated with lower energy consumption;
however, power must also be measured alongside with performance to observe the
overall energy consumption of a task.
\todo{Cite~\cite{Ruan2016}}
\todo{Summarize}
\todo{Relate to our work}
\todo{maybe don't immediately compare against papers and make a
  summary at the end of why this work is different to save space}

\todo{Cite Felter for this later: They found that using UnionFS has a negative performance penalty than using the native filesystem. Given this paper's findings, we make the following hypothesis about Docker's impact on energy usage: most CPU and memory-bound loads on Docker will use roughly the same energy as on native Linux; that loads that hit storage will use significantly more energy than on native Linux alone; and, without special configuration, loads heavy on networking may use more energy than on native Linux.}

\section{Methodology}
\label{sec:methodology}

\begin{figure}[tbh]
  \includegraphics[width=\columnwidth]{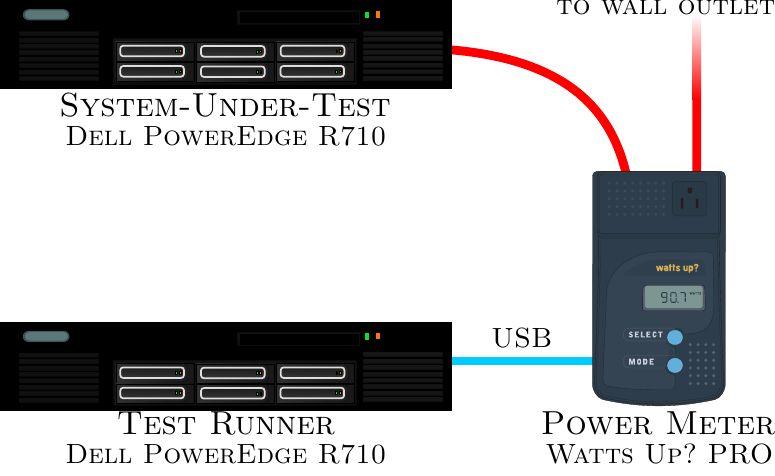}
  \caption{Hardware test setup: one rack-mount server
    System-Under-Test; and one test-runner. Power
    measurements were collected with a \wattsup{}.\label{fig:setup}
  }
\end{figure}

We want to compare the energy consumption of running a workload within a
Docker-managed container (the treatment) against running the same workload
on ``bare-metal'' (the control).
To estimate the energy consumption of one workload, we ran one server (the
\emph{system-under-test} or SUT) with the software of interest; we ran an
external system to initiate tests on the SUT and record the power measurements
(the \emph{test runner}); and we used a \emph{power meter} to measure the
instantaneous power consumed by the SUT.\@ We setup the systems to run the
desired software---either starting the service (bare-metal Linux) or start a
new container (Docker) that has already been built. We then initiated the tests on the test runner, which
would induce a workload on the SUT after a two minute pause. During the test
run, we collected root-mean-squared (RMS) power measurements, and recorded them.  We used the
power measurements to estimate the total energy consumption
on the SUT in two scenarios:\@ the software running on bare-metal Linux versus the
software running within a Docker container.

Importantly, the System-Under-Test is \textbf{not} the same machine as the
test runner; thus initiating the tests (test runner) is isolated from test execution (SUT).
Therefore, a separate server is used as the test runner for both
initiating tests and recording energy usage statistics from the power meter.

This section describes the hardware and instrumentation we used to run tasks
and collect power samples. An overview of our full setup is provided in
Figure~\ref{fig:setup}.  Our hardware setup consisted of a rack-mount server as
our System-Under-Test (Section~\ref{sec:sut}), a digital power meter
(Section~\ref{sec:wattsup}) to collect power samples, and a test runner
(Section~\ref{sec:runner}) to initiate the workloads.

\subsection{System-Under-Test}
\label{sec:sut}

The System-Under-Test (SUT) is a Dell PowerEdge R710 rack-mount server. A
summary of its hardware is listed in Table~\ref{tab:hardware}.  Although the
R710 is intended to be used with redundant power supplies, multiple network
interfaces, and redundant RAID storage, we only utilized one power supply, one
network interface (a gigabit Ethernet connection), and one hard drive for our tests.
The 2 Intel Xeon X5670s contain 6 cores each, totalling 12 real cores,
and with hyper-threading enabled they appear as 24 logical processors to Linux.

A summary of the software installed is listed in Table~\ref{tab:software}.
Docker was installed on the System-Under-Test. For bare-metal versions of Apache,
PHP, WordPress, MySQL, and PostgreSQL, we used \texttt{apt-\-get}.
Redis was installed from source on bare-metal Linux. All of the Docker application
software ran within Docker-managed containers. When installing software on Dock\-er, we used the official image hosted on Docker
Hub\cite{docker_hub_2016}. Note that the WordPress image
inherits from the \texttt{php:5.6-apache} image, which installs both PHP and
Apache. Hence, the only image we had to explicitly install was the one
containing WordPress.

\begin{table}[tbp]
\resizebox{\columnwidth}{!}{%
  \begin{tabular}{ll}
    \toprule
  CPU          & 2$\times$Six-core Intel Xeon X5670 at 2.93 GHz     \\
  RAM          & 72 GiB ECC DDR3                                    \\
  Network      & Gigabit Ethernet connection                        \\
  Storage      & 146GB SAS hard drive at 15000 RPM             \\
  Power supply & 870 Watts (120 volts $\sim$ 12A at 60 Hz)          \\
    \bottomrule
  \end{tabular}
}
\caption{Hardware configuration of the System-Under-Test and the test runner.}
\label{tab:hardware}
\end{table}

\begin{table}[tbp]
\resizebox{\columnwidth}{!}{%
  \begin{tabular}{lll}
    \toprule
  Software     & Version                   & Docker Image             \\
    \midrule
  Distribution & Ubuntu Server 16.04.1 LTS &                          \\
  Kernel       & Linux 4.4.0               &                          \\
  Docker       & 1.12.1                    &                          \\
  Apache       & 2.4.10                    & php:5.6-apache           \\
  PHP          & 5.6.24                    & php:5.6-apache           \\
  MySQL        & 5.7.15                    & mysql:5.7.15             \\
  WordPress    & 4.6.0                     & wordpress:4.6-apache     \\
  Redis        & 3.2.3                     & redis:3.2.3              \\
  PostgreSQL   & 9.5.4                     & postgres:9.5.4           \\
    \bottomrule
  \end{tabular}
}
\caption{Software versions used on the System-Under-Test}
\label{tab:software}
\end{table}

\subsection{Power measurements}
\label{sec:wattsup}

This paper focuses on comparing the energy required to perform several tasks.
However, we cannot measure energy directly. Instead, we measured the
instantaneous wall power drawn by the System-Under-Test. For this, we used a
\wattsup{}~\cite{wattsup_pro} power meter.

The \wattsup{} is a device with a Type B AC power socket. It samples the
voltage and current draw of the electrical appliance plugged into its socket.
Since power is voltage multiplied by current, the meter can report the
instantaneous power usage of an electrical appliance---in our case, a
rack-mount server as our System-Under-Test.  Since we are interested in the
total power usage of the entire system---including the CPU, but also memory,
storage, network interfaces, peripherals, internal cooling, and even overhead
due to the power supply---we opted to measure wall power, instead of using
onboard measurement, such as Intel's RAPL for measuring CPU power usage alone.
The \wattsup{} calculates the root-mean-square (RMS)\todo{explain why RMS?} of
thousands of samples over the course of one second~\cite{wattsup_vernier}.
Previous work by McCullough~\etal{mccullough_evaluating_2011} found that
collecting RMS measurements at a frequency of one measurement per second from
a \textsc{Watts Up?} power meter is sufficient for accurate energy consumption
estimation~\cite{mccullough_evaluating_2011}.

We used a modified version of \texttt{yyongpil}'s
\texttt{wattsup}\footnote{\url{https://github.com/yyongpil/wattsup}} software to
retrieve the power measurements from the \wattsup{} and save them on the
test runner. Every second, the wattage used by the System-Under-Test is pulled
from the \wattsup{}, transferred over USB to the test runner, and then written
to \texttt{std\-out}. Collection scripts on the test runner controlled the test runs for
each of our case studies and recorded measurements for each test run in order
to gather power data along with timestamps.  This information was saved to a
local SQLite3 database on the test runner.

\todo{Should I really explain why AC power is RMS averaged? I mean,
power is a function of current, which is ALTERNATING polarity}

However, power is not energy.  Energy is the integration of power over time.
The \wattsup{} yields RMS power samples of one second in
duration---several measurements of instantaneous power averaged over one
second.  Given an initial timestamp ($t_i$) and an end timestamp ($t_f$), we
can use the sum of power samples to estimate the energy required to complete a
task.  We approximated energy using a sum of power samples, taken at a regular
frequency. This is analogous to using the rectangle method of approximating an
integral with a duration $\Delta t$ of 1 second (Equation~\ref{eqn:energy}).

\begin{equation}\label{eqn:energy}
  E = \int_{t_i}^{t_f} P(t)\,dt
  \approx \Delta t \sum_{k=i}^{f} P_{RMS}(t_k)
\end{equation}

We wrote Python scripts that implemented the above estimation, taking in test
data from the SQL\-ite\-3 databases on the test runner, which had power in watts
with timestamps. Each timestamp was asserted to be about one second apart,
thus making our estimation valid. The summation produces an estimate of the
total energy consumed for a single run of a test. We considered each test run
to be one energy sample. We ran each test 40 times, giving us 40 energy
samples per case study per configuration.
\todo{Explain why multiple tests? Cite Green Mining, probably}
Before each test run, we had the machine sleep for two
minutes to reset the machine to its idle run state, as
Chowdhury~\etal{chowdhury_client-side_2016} discovered that running tests in
quick succession may alter the power state of the machine, artificially
skewing results. These energy summaries are then compared, grouped by case
study, for bare-metal Linux versus Docker.

\subsection{Test Runner}
\label{sec:runner}

For initiating the tests and recording the power samples, we used a Dell
PowerEdge R710 rack-mount server, identical in hardware specification and
configuration as the SUT.\@ We wrote collection scripts in Python that
initiate the tests (described in Section~\ref{sec:case-study}) on the
System-Under-Test through network requests, while simultaneously recording
energy statistics from the \wattsup{} via USB with yyongpil's
\texttt{wattsup}. We recorded timestamps for every power sample.

For each experiment:
\begin{enumerate}
  \tightlist%
  \item We started the service on the System-Under-Test (if applicable). In Docker,
    we started one or more new containers from their respective Docker images.
  \item On the test runner, we initiated a batch of test runs.
  \item For each test run, the test runner optionally performed a per-test initialization.
  \item The test runner would then sleep for two minutes.
  \item The test runner then induced a workload on the Sys\-tem-Under-Test via
    network requests.
  \item During each test run, the test runner recorded the instantaneous power
    measurements of the SUT and the timestamp every second.
  \item After all test runs from a batch have finished, we calculated the
    energy per each test run.
\end{enumerate}

The test runner was connected to the System-Under-Test via a gigabit switch.

\section{Case Study}
\label{sec:case-study}

Three open-source software projects were selected to test the difference in
energy consumption of running the app on bare-metal Linux versus within a
Docker-managed container.  Each of the applications stresses different
hardware resources, and together provides performance and energy insights on
which types of applications are most suited for Docker. WordPress with MySQL
represents an extremely popular website solution, while Redis and PostgreSQL
are common database solutions, with different use cases.  Considering the
popularity and breadth of applications selected as case studies, the results
give relevant insight into the effect of Docker on energy consumption when
compared with bare-metal installs.

\subsection{Idle}
\label{sec:idle}

As a baseline, we were interested in any possible overhead of running the
Docker service without placing any load on the system.  In order to estimate
how much energy is expected to be used at idle, the system was left to idle
for exactly 10 minutes, during which power usage was recorded.  In order to be
consistent with the methodology used for the following case studies, we inserted
an additional 2 minute of idle time before each test run during which power
samples were not recorded.  This test was performed 40 times sequentially, and
can be considered a baseline for bare-metal Linux and Docker.

``Idle'' means the system has been operating long enough to achieve a stable
state with nothing but the base operating system in operation, meaning that
none of the other services under test (PostgreSQL, Redis, MySQL, Apache) were
running, or were active in any way. When performing the Docker baseline, the
only difference is enabling the Docker background service. \textbf{Zero}
containers were running, so we measured the overhead of just the Docker daemon
itself.

Since time is fixed in this test, any difference in energy \emph{must} be due
to a difference in power consumption.

\subsection{WordPress}
\label{sec:wordpress}

\todo{Explain that DB folders are explicitly NOT mounted using AUFS!}

WordPress is an open-source content management system~\cite{wordpress}. As of
February 2017, Docker Hub has had over 10 million WordPress
pulls~\cite{docker_hub_2016} and WordPress powers over one quarter of the top
10 million websites worldwide~\cite{w3techs_2016}.

We installed WordPress manually for the bare-metal Linux version, as per the
WordPress official documentation~\cite{wordpress_install}. We used Docker
Compose~\cite{docker_compose_2016} for installing WordPress within Docker.
Both methods installed the same versions of WordPress, MySQL, PHP, and Apache,
as listed in Table~\ref{tab:software}.  On the bare metal system, MySQL and
Apache ran as services. Docker required two containers: one container held
Apache, which runs WordPress with \texttt{modphp}, while another contained
the MySQL database. These were automatically setup and connected using Docker
Compose. We generated a blog using the WP Example Content Plugin
1.3~\cite{wp_example_content_2016}, whose database was copied both into the
bare-metal installation and the Docker installation.

We used Tsung 1.6.0\cite{tsung} to perform an HTTP load stress test on the
WordPress server for which the test runner was monitoring energy usage. Tsung,
running on the test runner, created virtual clients that simulate a large
number of users visiting the WordPress front page and randomly navigate the
site. Each test was exactly 15 minutes long. Starting from no load, the test
added 100 simulated users per second. Each user loaded the WordPress homepage
content, which in turn required database queries in order to retrieve the
posts and other content.  We performed the full test 40 times sequentially, in
order to produce 40 energy samples, with 2 minutes of idle time between tests
to ensure accuracy of the energy measurements.

\subsection{Redis}
\label{sec:redis}

Redis is an open-source, in-memory key store that can be used as a database,
cache, or message broker~\cite{redis_2016}. As of February 2017, Docker Hub
has had over 10 million Redis pulls~\cite{docker_hub_2016}. We chose the Redis
to test the overhead of a workload that is predominantly memory, CPU, and
network bound (it does minimal accesses to storage).

Redis was installed in Docker with the version specified in
Table\ref{tab:software}. On bare-metal Linux, Redis was built from source. For Docker,
we used the official image to build a single container which held the Redis
server. The official image
downloaded from Docker Hub disables periodic persistence of the in-memory
database to permanent storage, hence we disabled this on the bare-metal
configuration as well.

The Redis benchmark suite, \texttt{redis-benchmark} was used to create a
workload of 1000 parallel clients making a total of 1.5 million requests. This
involves a great deal of network traffic from the server running the clients,
as well as doing a large amount of memory accesses.  We ran the full test 40 times sequentially, which
produced 40 energy samples, with two minutes of idle time between each sample.

\subsection{PostgreSQL}
\label{sec:postgres}

\todo{Explain that DB folders are explicitly NOT mounted using AUFS!}

PostgreSQL is an open-source, object-relational database management system
(DBMS)~\cite{postgresql}. As of February 2017, PostgreSQL has been pulled over
10 million times~\cite{docker_hub_2016}.

PostgreSQL includes \texttt{pgbench} for performance benchmarking.  PostgreSQL
was installed on both the SUT and test runner servers with the version
specified in Table~\ref{tab:software}. On bare-metal Linux, we ran PostgreSQL
as a service, while Docker held the database processes in a single container.
It is important to note that the Ubuntu 16.04 version enables SSL by default
whereas the Docker install does not. We accounted for this by disabling SSL in
the bare-metal Linux PostgreSQL installation. We also ran a test on the
bare-metal configuration with SSL enabled, to compare the overhead of Docker
against the overhead of encrypting queries. In Docker, the default PostgreSQL
image creates a volume mounted on the host (i.e., escaping the container) for
persisting data.  Thus, writes do not access Docker's AUFS storage layer.

The test consisted of running \texttt{pgbench} on the test runner with 50
clients, each peforming 1000 database transactions on the SUT of ``a scenario
that is loosely based on TPC-B''~\cite{pgbench,TPC-B}. We performed 40
sequential tests to produce 40 energy samples. Before each test, we ran
\texttt{pgbench -i} to initialize the database, then waited for two minutes of
idle time before starting the test proper. The entire test was performed for
both bare-metal Linux and Docker.

\section{Results}
\label{sec:results}

\newcommand{\EffectSize}{\multicolumn{2}{c}{Effect Size}}
\newcommand{\Correlation}{\multicolumn{2}{c}{Correlation ($r_{Et}$)}}

\begin{table}[tbp]
\resizebox{\columnwidth}{!}{%
  \begin{tabular}{ ll *{4}{r} }
\toprule
    Case Study & Normal &   \EffectSize\            &   \Correlation\  \\
                          \cmidrule(r{0.25em}){3-4}   \cmidrule(l{0.25em}){5-6}
               &        & Cliff's $d$ & Cohen's $d$ &   Linux & Docker \\
\midrule
    Idle       & No     &        0.80 &             &         &        \\
    WordPress  & No     &        1.00 &             &    0.83 &   0.99 \\
    Redis      & Yes    &             &      11.31  &    0.98 &   0.98 \\
    PostgreSQL & Yes    &             &       1.55  &    0.99 &   0.95 \\
\bottomrule
  \end{tabular}
}
  \caption{%
    Summary of results obtained for each experiment. ``Correlation'' refers to
    the linear correlation between estimated energy with the elapsed time of
    the test run.  Note: for the ``idle'' experiment, calculating correlation
    of energy with run time does not make sense because the elapsed time is
    fixed.
  }\label{tab:results}
\end{table}

After collecting all power samples, estimating energy per each test run, we
ran some statistical analyses on the results to determine whether there is a
significant difference in energy consumption to run a task on bare-metal Linux
compared to a Docker container. A summary of our results is given in
Table~\ref{tab:results}. Our raw data is available online.\footnote{%
  Available: \url{https://archive.org/details/docker-linux-energy-feb-2017.sqlite3}
}

First, we determined whether both energy samples on Linux and on Docker were
normally distributed using the Shapiro-Wilk normality test. Then, we applied
various tests to determine if both samples came from the same distribution.
For normally-distributed data, we used a paired Student's $t$-test.
Otherwise, we applied non-parametric tests: a Kruskal-Wallis rank sum test,
and a pairwise Wilcoxon rank sum test. In all two sample experiments,  we
found that the difference in distributions of energy consumption in Docker
compared with Linux was statistically significant, with a $p$-value near
zero,\footnote{%
  If the $p$-value is less than $10^{-4}$ (and thus, was only expressed using
  exponential notation), then we considered it to be ``near zero''.
} no matter which test we used. To quantify the difference, we calculated the
effect size. For all tests, we used Cliff's delta, which simply compares how
often samples from one distribution are greater than samples in the other
distribution. As shown in Table~\ref{tab:results}, for the WordPress and Redis
experiments, the distributions from Docker are all greater than the observations
from Linux with a maximum Cliff's delta of $1.0$. The other two experiments
also had large effect size, according to Cliff's delta, with small overlaps in
distributions.  Finally, we calculated the linear correlation,
Pearson's correlation coefficient, of energy with run
time. Recall that energy is $\text{power} \times \text{time}$.  Thus, energy
should be strongly correlated with time (an $r$ value of $+1.0$). In every
case, we found that energy was strongly correlated with time, however, since
the $r$ value of each test was not exactly $1.0$, we assert that other factors
must be influencing the total energy rather than energy being
completely explained by run time.

\todo{Explain how we will present the data: explain Violin plots, density
plots. What are the lines?}

The results are presented in two ways: summaries of the energy data is
presented in \textbf{violin plots} (Figures \ref{fig:idle-energy}, \ref{fig:wordpress-energy}, \ref{fig:redis-energy}, \ref{fig:postgres-energy}) 
which can read somewhat like box plots where each ``violin'' represents one
distribution. The width of the violin at any given point represents the
density of measurements observed at that point.  To give a sense of tendency,
a line is drawn at the median of the sample distribution.
Summaries of the power data are given as \textbf{density plots} (Figures \ref{fig:idle-power}, \ref{fig:wordpress-power}, \ref{fig:redis-power}, \ref{fig:postgres-power}), 
with hexagonal bins. Each bin represents a cluster of observations at the
given time and wattage.  Darker hexagons represent a denser concentration of
observations.

\subsection{Idle}

\begin{figure}[tbp]
\centering
\includegraphics[width=\columnwidth]{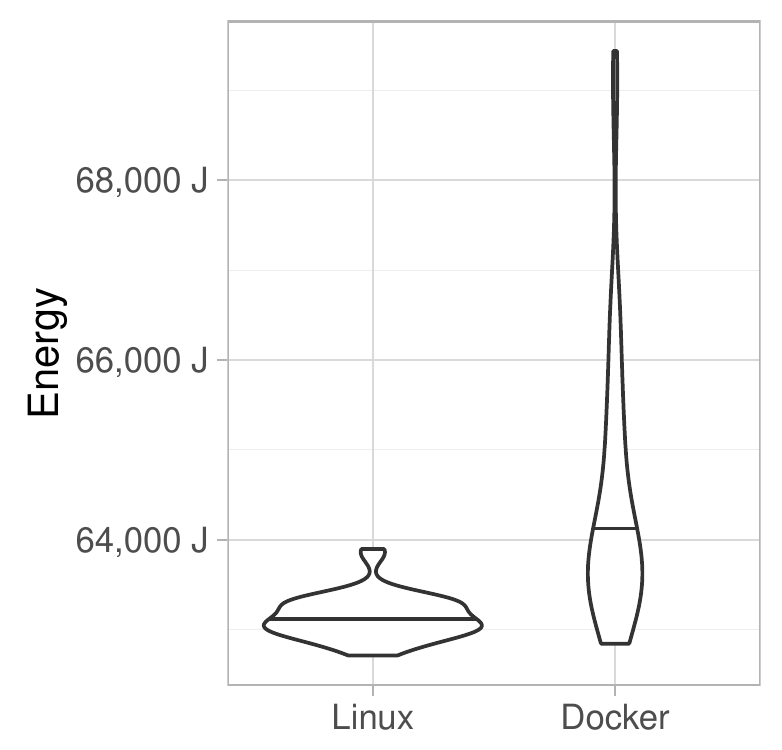}
\caption{Violin plot of idle energy consumption}
\label{fig:idle-energy}
\end{figure}

\begin{figure*}[tbp]
\centering
\includegraphics[width=\textwidth]{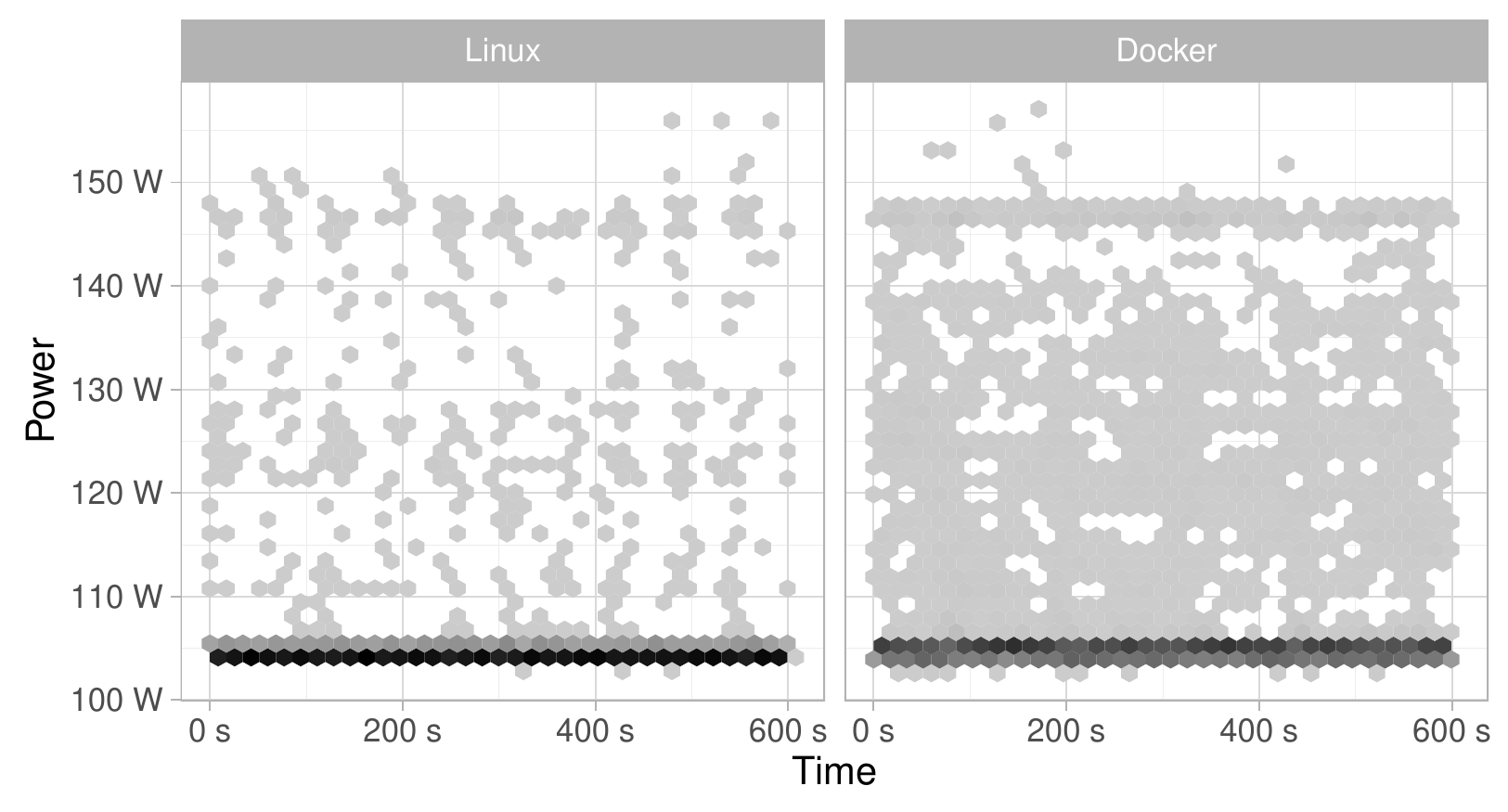}
\caption{Density plot of wattage measurements across all idle test runs over
time.}
\label{fig:idle-power}
\end{figure*}

\todo{Power plots: give standard deviation of Linux and docker measures}

The distribution of energy consumption with no load for 10 minutes is given as
a violin plot in Figure~\ref{fig:idle-energy}. A density of power is provided
in Figure~\ref{fig:idle-power}.  Using the Shapiro-Wilk test, we found that
neither the bare-metal Linux nor the Docker distributions are
normally-distributed. Using the non-parametric Kruskal-Wallis test and
pairwise Wilcoxon rank sum test, we obtained a $p$-value close to zero,
indicating that the distributions are indeed different. Using Cliff's delta,
we got an effect size of $0.80$, indicating that values in the Docker
distribution are nearly $80\%$ likely to be greater than an observation in the
bare-metal Linux distribution. Another way to think about this difference, is
that three-quarters of the time, we observed that running on bare-metal Linux
with no load would use less than 63,380 joules of energy, whereas if simply
the Docker daemon was running (with no containers running), three-quarters of
time we would observe the machine consuming \emph{more than} 63,380 joules of
energy for doing \emph{nothing} for ten minutes. This energy difference cannot
be attributed to performance, since time is fixed to 10 minutes in both cases.

This baseline establishes that, since the Docker daemon is an unavoidable
service that must run---regardless if containers are running or not---running
Docker comes with a power overhead. Whether this difference in energy
consumption over time is negligible is for operators to decide, however, later
we describe how to make back the difference in energy consumption.

\todo{%
  Say we can regain the energy lost here in other scenarios such as the
  PostgreSQL case.
}

\todo{How much money will I spend per month simply running the Docker daemon?
  Compare this to leaving on a 40W light bulb for the same amount of time.
}

\subsection{WordPress}

\begin{figure}[tbp]
\centering
\includegraphics[width=\columnwidth]{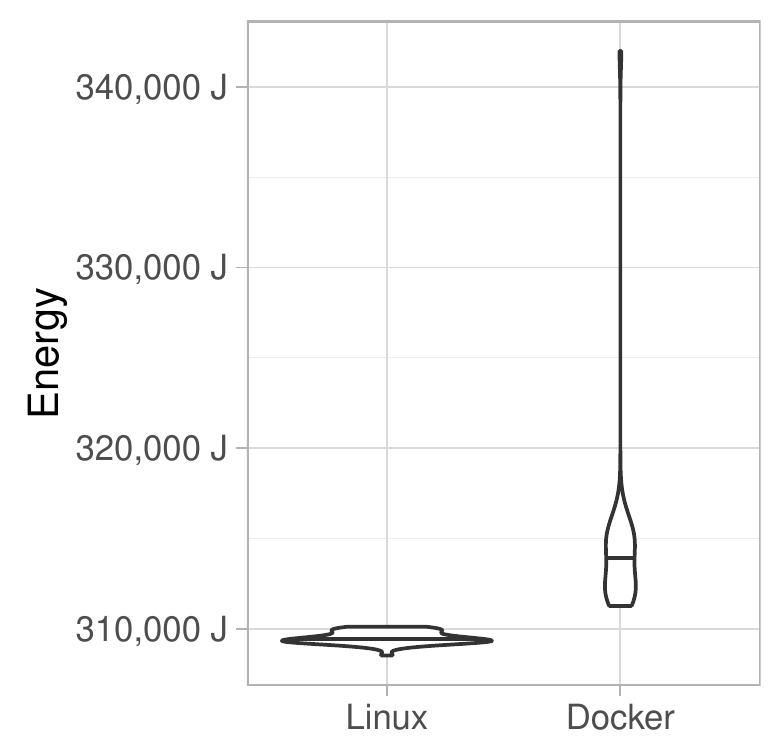}
\caption{Violin plot of energy consumption in the WordPress experiment}
\label{fig:wordpress-energy}
\end{figure}

\begin{figure*}[tbp]
\centering
\includegraphics[width=\textwidth]{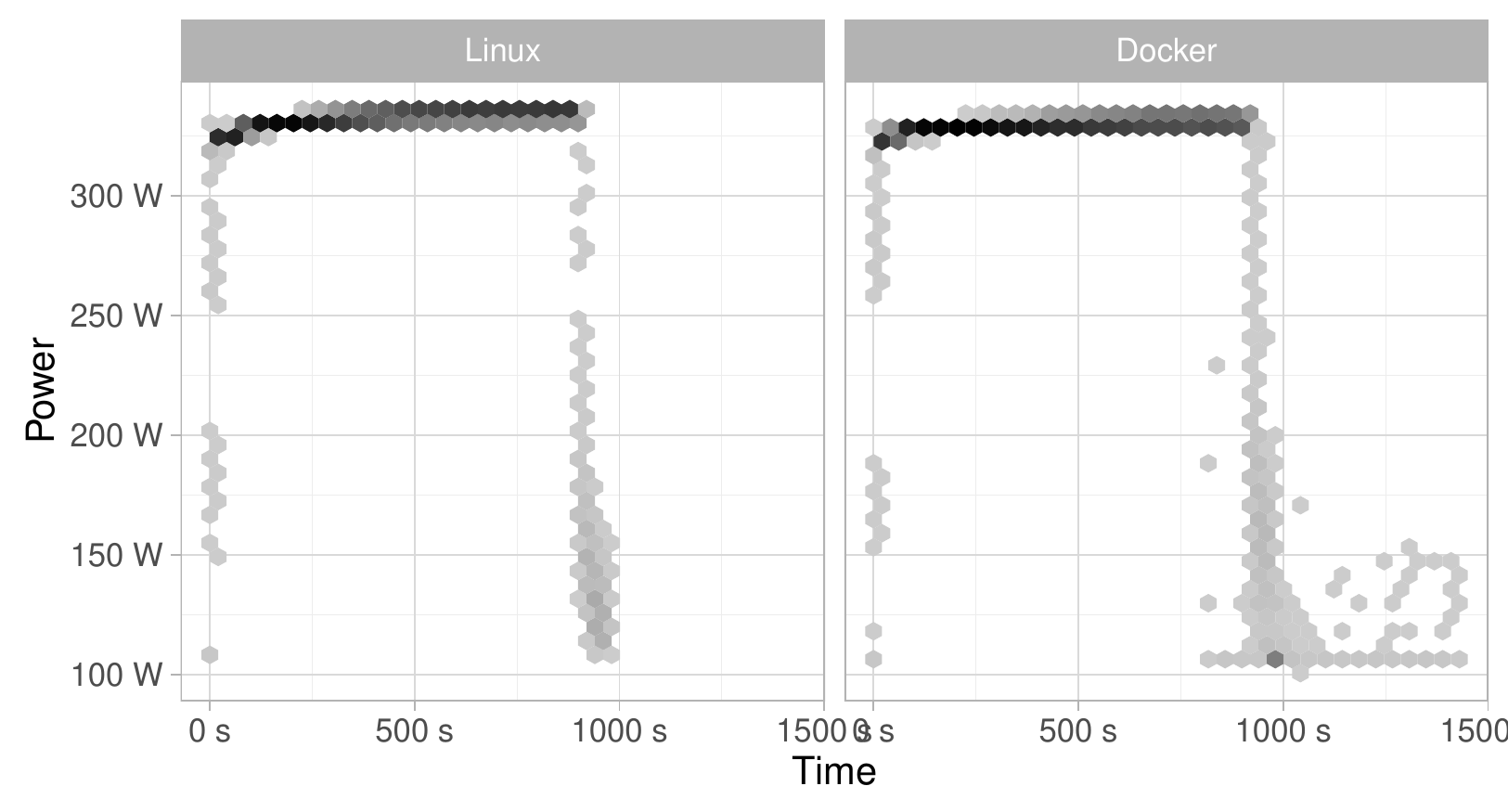}
\caption{Density plot of wattage measurements across all WordPress runs over
time.}
\label{fig:wordpress-power}
\end{figure*}

The distribution of energy consumption for running a simulated load on a
WordPress server under Linux and within Docker is shown in
Figure~\ref{fig:wordpress-energy}. A density of power is provided in
Figure~\ref{fig:wordpress-power}.  Using the Shapiro-Wilk test, only the
distribution of energy consumption under bare-metal Linux was
normally-distributed; hence, we used non-parametric tests for comparison and
effect size. Both the Kruskal-Walis and pairwise Wilcoxon rank sum test
yielded a $p$-value near zero, meaning that the distributions are
significantly different. For effect size, we computed a Cliff's delta of
$1.0$, implying completely non-overlapping distributions. In other works,
\emph{all} samples in the Docker test runs were higher than all samples in
bare-metal Linux. Finally, the linear correlation of energy and run time for
bare-metal Linux and Docker were of $0.8303$ and $0.9885$ respectively.

\subsection{PostgreSQL}

\begin{figure}[tbp]
\centering
\includegraphics[width=\columnwidth]{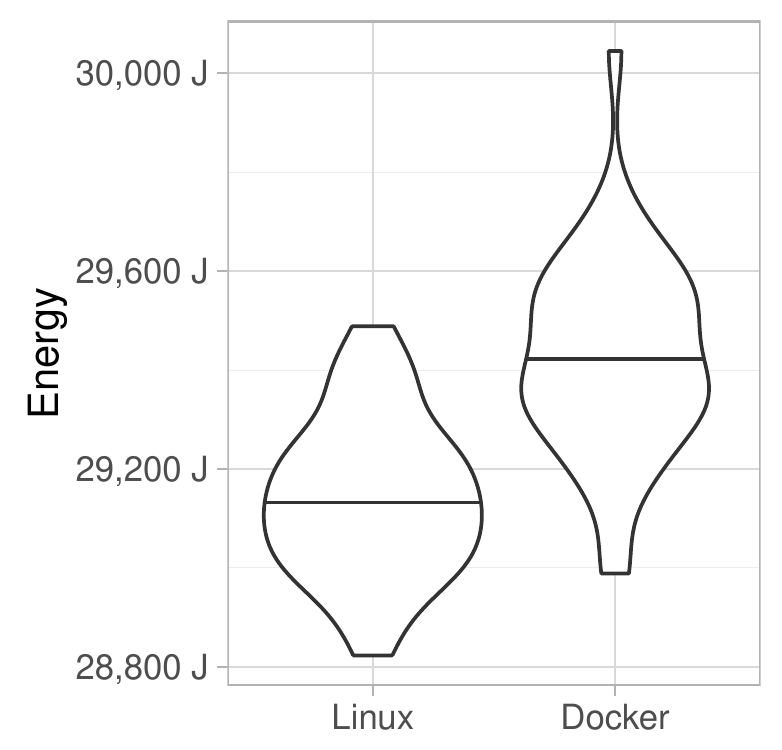}
\caption{Violin plot of energy consumption in the PostgreSQL test}
\label{fig:postgres-energy}
\end{figure}

\begin{figure*}[tbp]
\centering
\includegraphics[width=\textwidth]{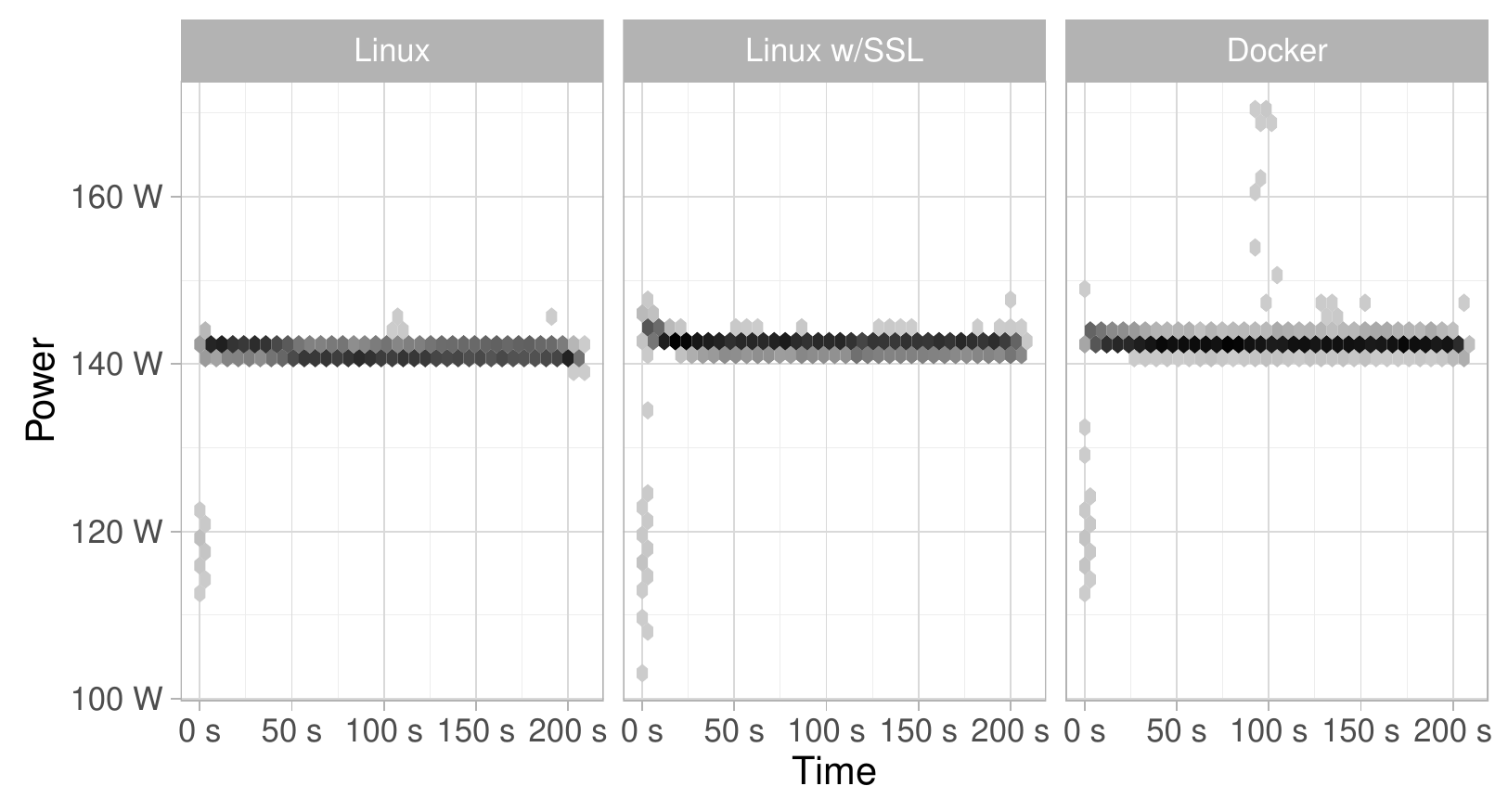}
\caption{Density plot of wattage measurements across all \texttt{pgbench} runs
over time.}
\label{fig:postgres-power}
\end{figure*}

For this test, we had three energy consumption distributions: bare-metal Linux,
with SSL disabled; bare-metal Linux, with SSL enabled; and Docker, with SSL
disabled. Not only are we testing the difference between Docker and
bare-metal, but we are also introducing the difference between encrypting
connections on bare-metal as well.
The three energy distributions are shown in
Figure~\ref{fig:postgres-energy}. A density of power is provided in
Figure~\ref{fig:postgres-power}. Using the Shapiro-Wilk test, we determined
that all three samples are normally distributed, with the smallest $p$-value
being $0.54$ for the Docker energy distribution. Thus, we used pairwise
paired Student's $t$-tests to compare each distribution to the others.
The baseline (Linux with SSL disabled) is significantly different, both to
Docker with SSL disabled, and with Linux with SSL enabled, with $p$-values
near zero. Interestingly, Docker with SSL disabled is \textbf{not} significantly
different compared to Linux with SSL enabled, with a $p$-value of 0.15. This
implies that the trade-off between encrypting connections with SSL is similar
to the trade-off between using Docker without encryption.

To understand the effect size, we used Cohen's $d$. Cohen's $d$
compares the means of the two normally-distributed samples, taking in to
account their pooled standard deviation to determine the
offset~\cite{cohens_d}. Larger results indicate a larger difference in the means.
Comparing PostgreSQL with SSL disabled on bare-metal Linux versus
the same configuration in Docker yields a very large Cohen's $d$ of $1.55$.  However,
\todo{Elaborate on this:}
simply turning on SSL on bare-metal Linux, testing against Docker with SSL
disabled yields the smallest effect size obtained in this paper: $0.31$.
This corroborates the findings of Chowdhury
\etal{chowdhury_client-side_2016} that simply using SSL/TLS
has a significant effect on energy consumption. The difference
between bare-metal Linux versus enabling SSL on the same configuration also has
a large effect size, with Cohen's $d$ calculated to be $1.32$.

\subsection{Redis}

\begin{figure}[tbp]
\centering
\includegraphics[width=\columnwidth]{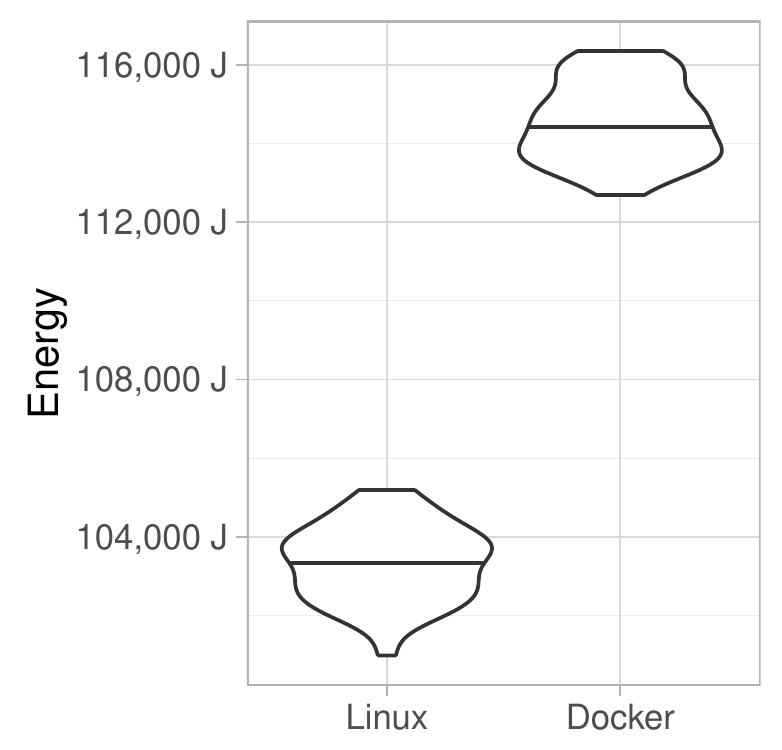}
\caption{Violin plot of energy consumption running the Redis benchmark.}
\label{fig:redis-energy}
\end{figure}

\begin{figure}[tbp]
\centering
\includegraphics[width=\columnwidth]{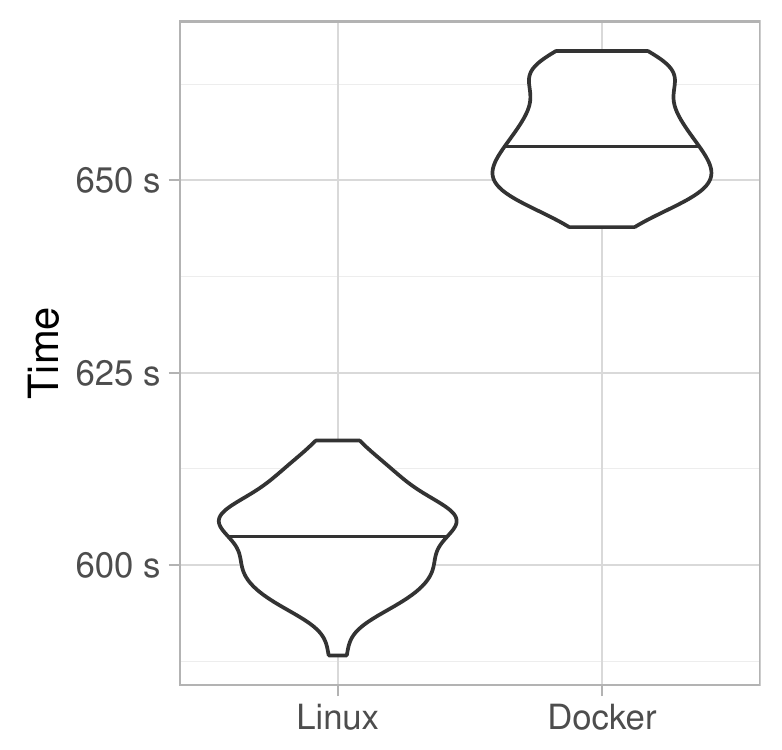}
\caption{Violin plot of elapsed time running the Redis benchmark.
  Elapsed time may explain the difference in energy (Figure~\ref{fig:redis-energy})
}
\label{fig:redis-time}
\end{figure}

\begin{figure*}[tbp]
\centering
\includegraphics[width=\textwidth]{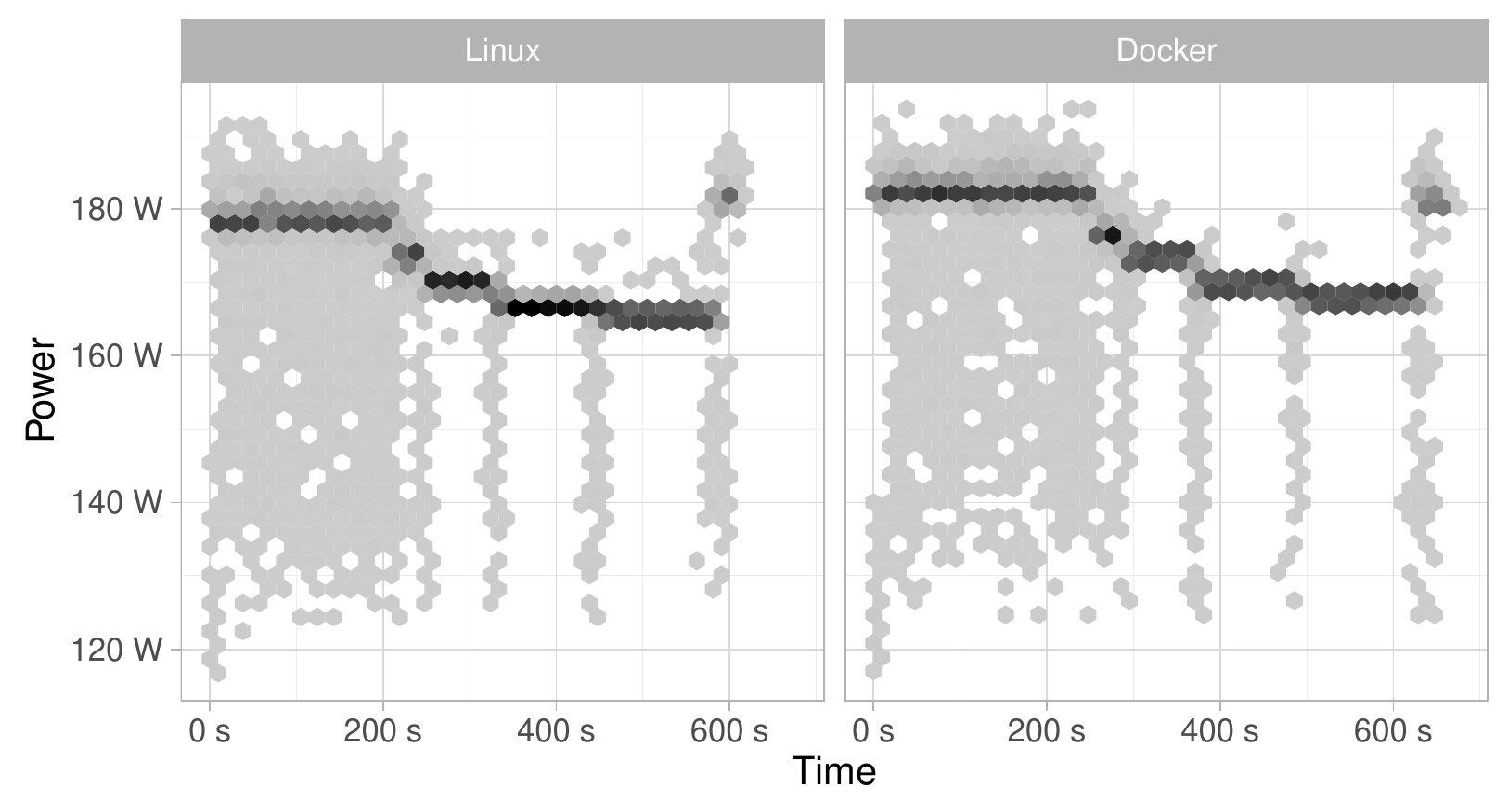}
\caption{Density plot of wattage measurements across all Redis Benchmark runs over
time.}
\label{fig:redis-power}
\end{figure*}

The distribution of energy consumption for running \texttt{re\-dis-\-bench} on Linux
and within Docker is shown in Figure~\ref{fig:redis-energy}. A density of
power is provided in Figure~\ref{fig:redis-power}. Using the Shapiro-Wilk
test, both samples are normally-distributed. We compared
the distributions using a Student's $t$-test and obtained $p$-values near
zero. Using Cohen's $d$, we obtained a huge effect
size of $11.31$. Thus, this experiment shows the greatest difference between running in Docker
versus running on bare-metal Linux.

The linear correlation of energy with time yielded $0.996$ and $0.995$ for
Linux and Docker, respectively.  Given the very high correlation of energy
with time, we also compared the amount of time it took to complete each test
(Figure~\ref{fig:redis-time}). Since elapsed time is greater under Docker,
energy consumption will be greater, unless the power used in Docker is
drastically lower, which is not the case (Figure~\ref{fig:redis-power}).

\section{Discussion}
\label{sec:discussion}

\newcommand{\unixwrite}{\texttt{write()}} 

    \todo{Concretely express the difference between Docker and not Docker in
    kilowatt hours, or even in cash money. Talk about monthly operating costs.}
    \todo{Mention how Docker uses crazy AUFS mounts to support writing to a
    read-only file system.}
    \todo{Cite~\cite{docker-aufs}: ``the AUFS storage driver can introduce
    significant latencies into container write performance''. This is due to COW.}
    \todo{Also from~\cite{docker-aufs}: ``Data volumes provide the best and most
    predictable performance. This is because they bypass the storage driver and do
    not incur any of the potential overheads introduced by thin provisioning and
    copy-on-write''.}

Figure~\ref{fig:idle-energy} shows that having the Docker service running
consumes significantly more energy at idle than without Docker.  The
\texttt{dockerd} background process explains the difference in energy
consumption. Recall that Docker is \emph{not} required for containerization;
rather, Docker provides a convenient infrastructure for running containerized
applications in Linux.  However, \texttt{dockerd}, the Docker server,
written in the Go programming language periodically wakes up to do work, even
if it is managing zero active containers. Using \texttt{perf top -p \$(pgrep dockerd)}
we found that the \texttt{dockerd} was periodically calling functions related
to scheduling and garbage collection in Go (e.g.,
\texttt{run\-time.find\-runnable}, \texttt{run\-time.scan\-object},
\texttt{run\-time.heap\-Bi\-ts\-For\-Object}, \texttt{run\-time.grey\-object}).

\todo{explain WordPress}

A possible service deployment strategy is to create virtual networks wherein
each microservice is in its own container. Only public-facing services (of
which there should be few) will be required to use any kind of
per-connection encryption, as provided by SSL/TLS\@. Our results show that,
while PostgreSQL in Docker uses more energy compared to the same configuration
in Linux, the effect is not very large compared with running PostgreSQL on
Linux with encryption turned on. In that case, running PostgreSQL within
containers, with unencrypted inter-container communication may actually be
a more energy efficient option.

\todo{Redo Redis strace}

Using \texttt{strace -c}, we measured the time spent in system calls running
the \texttt{redis-bench} application. We found that in both bare-metal Linux and
in Docker, the Redis server was mostly calling  \unixwrite\ (about 82\%
of all system calls). A 32--39 second benchmark induced around 1.7 million
\unixwrite\ system calls. The notable difference is that the Redis server
running within a Docker container spent more than twice as long doing writes
(93.94 milliseconds) versus running the server on bare-metal Linux (44.08
milliseconds spent in \unixwrite). This explains a small part of the
longer runtime on Docker (and thus higher energy consumption), though it does
not come close to explaining the large gap in run time.

\section{Threats to validity}
\label{sec:tov}


\paragraph{Construct validity} In general, using benchmark frameworks does not
necessarily model real usage of the applications. This is especially true when
there has been no investigation in to what a realistic typical usage of these
applications would be, as was the case here. Future work should start by
discovering what is representative of typical usage for each of the test cases
(a profile), or benchmark using real world data and actions, if at all
possible.

Docker has a number of configuration options concerning networking and the
file system. Likely, any administrator deploying Docker in production would
tweak these settings extensively. As such, our usage of ``off-the-shelf'' defaults
(deploying straight from the Docker Hub image using a command like \texttt{docker run postgres:latest})
is not representative of true deployments using Docker.

Each of the studied applications was only serving a single host.  Each
benchmarking tool provided support for simulating multiple clients, and these
features were used in all tests. The quality of the multiple simulated clients
from a single client when compared to real-world users is unknown and so may not
realistically stress the applications. Furthermore, the servers were only using a
single gigabit Ethernet connection, where real deployments may see multiple
network connections sharing the load of requests.

\paragraph{Internal validity} One may call into question the precision and
reliability of the power measurements obtained from the \wattsup{}. Another
threat to validity is that we left services such as \texttt{OpenSSH} and
\texttt{OpenVPN} running on the System-Under-Test, whose power usage is also
included in all of the power measurements. Thus, the \emph{exact} numbers may
not be indicative of real loads, but, after taking several energy samples, the
comparisons can give an idea of the differences. SSH and VPN were only used
for configuring the machines before the tests were run; none of the traffic in
any of the tests used SSH or used the same network interface as the VPN\@.

\paragraph{External validity} The applications selected as test cases do not
necessarily apply to other applications, even of similar type. Generalizations
are hard to draw from such a small set of applications. Even different
versions of the same application have different energy
profiles~\cite{energy_versions,greenoracle}---especially when the load makes
different operating system calls.  External parties need to consider the
resources required by their application in order to best evaluate the
consequences of using Docker.

Finally, the System-Under-Test that we used only represents a single machine
configuration. Having multiple test platforms that differ in performance and
architecture would allow for more generalized findings.

\section{Conclusion}
\label{sec:conclusion}


In this paper, we compared the energy consumption of various workloads running
within Docker-managed containers and on ``bare-metal'' Linux.
After almost 2 days and 20 hours of total time collecting power measurements, we
found that, in all cases, workloads running in Docker have a measurable
energy overhead.
Simply running \texttt{dockerd} idle induces a 2 watt difference in average
power, and thus an increase in energy over time.
However, the increase in energy consumption may mostly be attributed
to runtime
performance. In the case of Redis and WordPress, the increase in
energy can be attributed to increase in runtime---thus the decrease in
performance explains the increase in total energy consumption.

Operations teams must decide which is more important: sustainability
and energy consumption and run-time performance of reduced resource usage by employing
bare-metal Linux, or the process isolation and maintainability of
containerized applications of Docker. Saving on heat and energy is
important for some scenarios, yet the human cost of maintenance can
far exceed run time, energy, and heating costs of Docker's minor inefficiency.

\todo{%
    make this sentence better: It is up to the operations team to judged
    whether easier deployment is worth the energy and performance overhead
}

\FloatBarrier%
\pagebreak[3]
\bibliographystyle{acm}
\bibliography{bib/bibliography,bib/autogenerated}

\begin{thebibliography}{10}

\bibitem{docker_adoption}
{\sc Arijs, P.}
\newblock Docker usage statistics: Increased adoption by enterprises and for
  production use.
\newblock
  \href{http://www.coscale.com/blog/docker-usage-statistics-increased-adoption-by-enterprises-and-for-production-use}{http://www.coscale.com/blog/docker-usage-statistics-increased-adoption-by-enterprises-and-for-production-use},
  July 2016.
\newblock (Accessed on 12/21/2016).

\bibitem{LXC}
{\sc {Canonical Ltd.}}
\newblock Linux containers.
\newblock \url{https://linuxcontainers.org/}, February 2017.
\newblock (Accessed on 02/06/2017).

\bibitem{greenoracle}
{\sc Chowdhury, S.~A., and Hindle, A.}
\newblock {GreenOracle}: Estimating software energy consumption with energy
  measurement corpora.
\newblock In {\em Proceedings of the 13th International Conference on Mining
  Software Repositories\/} (New York, NY, USA, 2016), MSR '16, ACM, pp.~49--60.

\bibitem{chowdhury_client-side_2016}
{\sc Chowdhury, S.~A., Sapra, V., and Hindle, A.}
\newblock {C}lient-{Side} {Energy} {Efficiency} of {H{T}TP}/2 for {Web} and
  {Mobile} {App} {Developers}.
\newblock In {\em 2016 {IEEE} 23rd {International} {Conference} on {Software}
  {Analysis}, {Evolution}, and {Reengineering} ({SANER})\/} (March 2016),
  vol.~1, pp.~529--540.

\bibitem{cohens_d}
{\sc Cohen, J.}
\newblock Statistical power analysis for the behavioural sciences. hillside.
\newblock {\em NJ: Lawrence Earlbaum Associates\/} (1988).

\bibitem{rkt}
{\sc {{CoreOS}, {Inc.}}}
\newblock {rkt}, a security-minded, standards-based container engine.
\newblock \url{https://coreos.com/rkt}, February 2017.
\newblock (Accessed on 02/06/2017).

\bibitem{docker_compose_2016}
Overview of {D}ocker {C}ompose.
\newblock \url{https://docs.docker.com/compose/overview/}, September 2016.
\newblock Accessed: 2016-09-02.

\bibitem{docker_hub_2016}
Docker {H}ub.
\newblock \url{https://hub.docker.com/explore/}, September 2016.
\newblock Accessed: 2016-09-02.

\bibitem{what_docker}
{\sc {Docker Inc.}}
\newblock What is docker?
\newblock \url{https://www.docker.com/what-docker#/VM}, November 2016.
\newblock (Accessed on 12/21/2016).

\bibitem{ellis_case_1999}
{\sc Ellis, C.~S.}
\newblock {T}he case for higher-level power management.
\newblock In {\em Hot {Topics} in {Operating} {Systems}, 1999. {Proceedings} of
  the {Seventh} {Workshop} on\/} (1999), IEEE, pp.~162--167.

\bibitem{felter_updated_2015}
{\sc Felter, W., Ferreira, A., Rajamony, R., and Rubio, J.}
\newblock {A}n updated performance comparison of virtual machines and linux
  containers.
\newblock In {\em Performance {Analysis} of {Systems} and {Software}
  ({ISPASS}), 2015 {IEEE} {International} {Symposium} {On}\/} (2015), IEEE,
  pp.~171--172.

\bibitem{clusterhq_survey}
{\sc Ferranti, M.}
\newblock Survey: 96\% increase in container production usage over past year ·
  clusterhq.
\newblock \url{https://clusterhq.com/2016/06/16/container-survey/}, June 2016.
\newblock (Accessed on 01/30/2017).

\bibitem{wp_example_content_2016}
{\sc Ferrara, J.}
\newblock {WP} {E}xample {C}ontent.
\newblock \url{https://wordpress.org/plugins/wp-example-content/}, September
  2016.

\bibitem{gupta_detecting_2011}
{\sc Gupta, A., Zimmermann, T., Bird, C., Nagappan, N., Bhat, T., and Emran,
  S.}
\newblock {D}etecting {Energy} {Patterns} in {Software} {Development}.
\newblock {\em Microsoft Research Microsoft Corporation One Microsoft Way
  Redmond, WA 98052\/} (2011).

\bibitem{hindle_green_2012}
{\sc Hindle, A.}
\newblock {G}reen mining: {A} methodology of relating software change to power
  consumption.
\newblock IEEE, pp.~78--87.

\bibitem{KVM}
{\sc {{Linux} Kernel Organization, {Inc.}}}
\newblock Kvm.
\newblock \url{http://www.linux-kvm.org/page/Main_Page}, November 2016.
\newblock (Accessed on 02/06/2017).

\bibitem{mccullough_evaluating_2011}
{\sc McCullough, J.~C., Agarwal, Y., Chandrashekar, J., Kuppuswamy, S.,
  Snoeren, A.~C., and Gupta, R.~K.}
\newblock {E}valuating the effectiveness of model-based power characterization.
\newblock In {\em {USENIX} {Annual} {Technical} {Conf}\/} (2011), vol.~20.

\bibitem{morabito_power_2015}
{\sc Morabito, R.}
\newblock {P}ower {Consumption} of {Virtualization} {Technologies}: an
  {Empirical} {Investigation}.
\newblock {\em arXiv preprint arXiv:1511.01232\/} (2015).

\bibitem{postgresql}
{About}---{PostgreSQL}.
\newblock \url{https://www.postgresql.org/about/}, 2016.
\newblock Accessed: 2016-09-09.

\bibitem{pgbench}
{\sc The {PostgreSQL Global Development Group}}.
\newblock {\em {pgbench(1)}}, {PostgreSQL} 9.5.4~ed., 2016.

\bibitem{redis_2016}
Redis.
\newblock \url{http://redis.io/}, September 2016.
\newblock Accessed: 2016-09-02.

\bibitem{energy_versions}
{\sc Romansky, S., and Hindle, A.}
\newblock On improving green mining for energy-aware software analysis.
\newblock In {\em Proceedings of 24th Annual International Conference on
  Computer Science and Software Engineering\/} (Riverton, NJ, USA, 2014),
  CASCON '14, IBM Corp., pp.~234--245.

\bibitem{shea_power_2014}
{\sc Shea, R., Wang, H., and Liu, J.}
\newblock {P}ower consumption of virtual machines with network transactions:
  {Measurement} and improvements.
\newblock In {\em {IEEE} {INFOCOM} 2014 - {IEEE} {Conference} on {Computer}
  {Communications}\/} (apr 2014), pp.~1051--1059.

\bibitem{TPC-B}
{\sc {TPC}}.
\newblock {TPC-B}.
\newblock \url{http://www.tpc.org/tpcb/}, 1990.
\newblock (Accessed on 01/25/2017).

\bibitem{tsung}
Tsung.
\newblock \url{http://tsung.erlang-projects.org/}, 2016.
\newblock Accessed: 2016-09-07.

\bibitem{van2016power}
{\sc van Kessel, J., Taal, A., and Grosso, P.}
\newblock Power efficiency of hypervisor-based virtualization versus
  container-based virtualization.
\newblock {\em University of Amsterdam\/} (2016).

\bibitem{vasic_making_2009}
{\sc Vasi{\' c}, N., Barisits, M., Salzgeber, V., and Kostic, D.}
\newblock {M}aking {Cluster} {Applications} {Energy}-aware.
\newblock In {\em Proceedings of the 1st {Workshop} on {Automated} {Control}
  for {Datacenters} and {Clouds}\/} (New York, NY, USA, 2009), {ACDC}
  \textquotesingle 09, ACM, pp.~37--42.

\bibitem{w3techs_2016}
{\sc W3Techs.com}.
\newblock Usage of content management systems for websites.
\newblock
  \url{https://w3techs.com/technologies/overview/content_management/all/},
  September 2016.
\newblock Accessed: 2016-09-02.

\bibitem{wattsup_pro}
{Watts Up?} plug load meters.
\newblock
  \url{https://www.wattsupmeters.com/secure/products.php?pn=0&wai=322&more=4},
  2016.
\newblock Accessed: 2016-09-09.

\bibitem{wattsup_vernier}
{Watts Up Pro}.
\newblock \url{http://www.vernier.com/products/sensors/wu-pro/}, 2016.
\newblock (Accessed on 12/21/2016).

\bibitem{wordpress}
{A}bout---{Word{P}ress}.
\newblock \url{https://wordpress.org/about/}, 2016.
\newblock Accessed: 2016-09-05.

\bibitem{wordpress_install}
Installing {WordPress}---{WordPress Codex}.
\newblock \url{https://codex.wordpress.org/Installing_WordPress}, 2016.
\newblock (Accessed on 12/22/2016).

\bibitem{xu_energy_2015}
{\sc Xu, C., Zhao, Z., Wang, H., Shea, R., and Liu, J.}
\newblock {E}nergy {Efficiency} of {Cloud} {Virtual} {Machines}: {From}
  {Traffic} {Pattern} and {C{P}U} {Affinity} {Perspectives}.
\newblock {\em IEEE Systems Journal PP}, 99 (2015), 1--11.

\bibitem{zhang_accurate_2010}
{\sc Zhang, L., Tiwana, B., Qian, Z., Wang, Z., Dick, R.~P., Mao, Z.~M., and
  Yang, L.}
\newblock {A}ccurate online power estimation and automatic battery behavior
  based power model generation for smartphones.
\newblock In {\em Proceedings of the eighth {IEEE}/{ACM}/{IFIP} international
  conference on {Hardware}/software codesign and system synthesis\/} (2010),
  ACM, pp.~105--114.

\end{thebibliography}

\end{document}